\documentclass[runningheads]{llncs}
\usepackage{graphicx}
\usepackage{booktabs}
\usepackage{multirow}
\usepackage{makecell}
\usepackage{rotating}
\usepackage{dingbat}
\usepackage{xcolor}
\usepackage{amsmath,bm}

\newcommand{\etal}{\emph{et al.}}
\newcommand{\ie}{\emph{i}.\emph{e}. }
\newcommand{\eg}{\emph{e}.\emph{g}. }
\begin{document}
\title{Coherence Learning using Keypoint-based Pooling Network for Accurately Assessing Radiographic Knee Osteoarthritis}
%
%
%
\author{
Kang Zheng\inst{1} \and
Yirui Wang\inst{1}  \and
Chen-I Hsieh\inst{2} \and
Le Lu\inst{1} \and
Jing Xiao\inst{2} \and \\
Chang-Fu Kuo\inst{2} \and
Shun Miao\inst{1}
}
%
\authorrunning{K. Zheng et al.}

\institute{PAII Inc., Bethesda, MD, USA \and
Chang Gung Memorial Hospital, Linkou, Taiwan, ROC \and
Ping An Technology, Shenzhen, China}
%
%
%
\maketitle              
\begin{abstract}

Knee osteoarthritis (OA) is a common degenerate joint disorder that affects a large population of elderly people  worldwide. Accurate radiographic assessment of knee OA severity plays a critical role in chronic patient management. Current clinically-adopted knee OA grading systems are observer subjective and suffer from inter-rater disagreements. In this work, we propose a computer-aided diagnosis approach to provide more accurate and consistent assessments of both composite and fine-grained OA grades simultaneously. A novel semi-supervised learning method is presented to exploit the underlying coherence in the composite and fine-grained OA grades by learning from unlabeled data. By representing the grade coherence using the log-probability of a pre-trained Gaussian Mixture Model, we formulate an incoherence loss to incorporate unlabeled data in training. The proposed method also describes a keypoint-based pooling network, where deep image features are pooled from the disease-targeted keypoints (extracted along the knee joint) to provide more aligned and pathologically informative feature representations, for accurate OA grade assessments. The proposed method is comprehensively evaluated on the public Osteoarthritis Initiative (OAI) data, a multi-center ten-year observational study on 4,796 subjects. Experimental results demonstrate that our method leads to significant improvements over previous strong whole image-based deep classification network baselines (like ResNet-50).

\keywords{Knee Osteoarthritis, Coherence Learning, Keypoint-based Pooling.}
\end{abstract}

\section{Introduction}
\begin{figure}[t]
\centering
\includegraphics[width=0.9\linewidth]{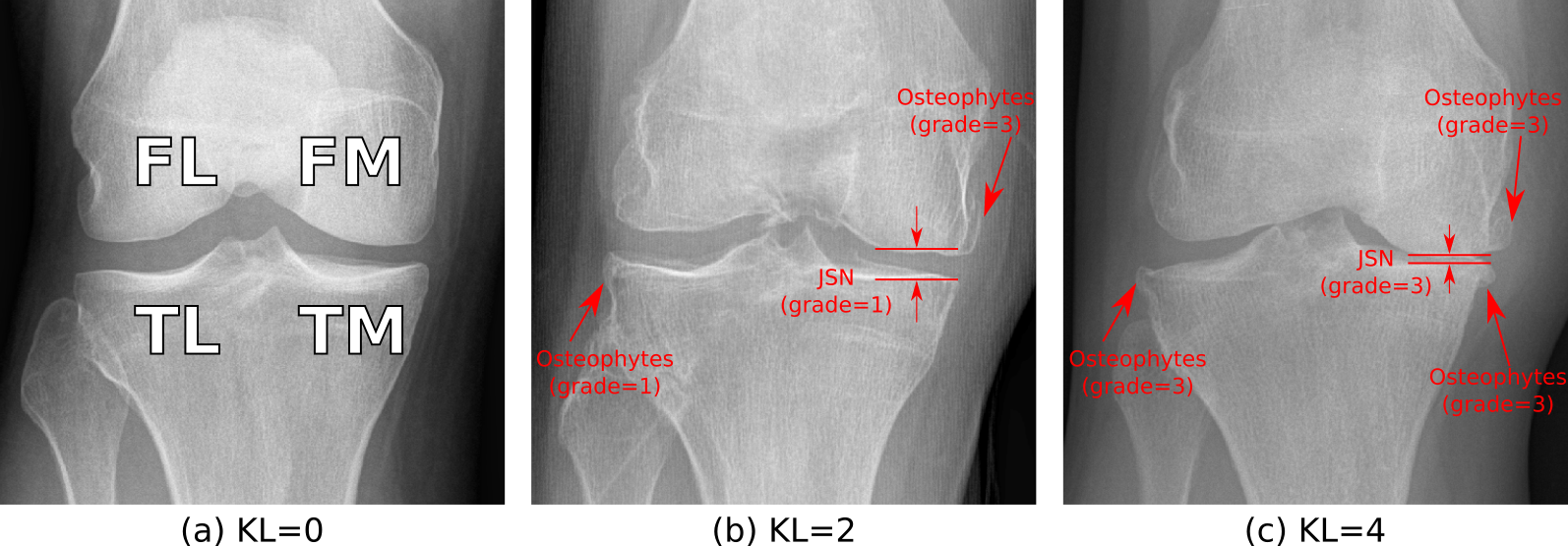}\vspace{-2mm}
\caption{Examples of knee OA with different KL, JSN and osteophyte grades. JSN is assessed separately for lateral and medial compartments. Osteophyte is assessed separately for femur lateral (FL), femur medial (FM), tibia lateral (TL), and tibia medial (TM) compartments.}
\label{fig:oarsi} \vspace{-5mm}
\end{figure}

Knee osteoarthritis (OA) is the most common joint disorder in the elderly and affects approximately 10\% men and 13\% women aged 60 years or older~\cite{zhang2010epidemiology}. It is also one of the most frequent causes of physical disability among older adults, which leads to significant social and economic burdens. Since there is no effective cure for OA, accurate assessment of OA severity is critical in the management of OA patients. Plain radiographs is an inexpensive tool for the diagnosis of OA~\cite{oka2008fully,shamir2008knee,shamir2009early}. However, quantifying OA severity accurately and objectively using plain ragiographs is a non-trivial task. The most widely adopted knee OA grading system in clinical practices is the Kellgren-Lawrence (KL) system~\cite{kellgren1957radiological}, as a semi-quantitative 5-scale assessment of the overall OA severity from the whole joint, by considering various factors including the joint space and osteophyte. Another grading system by Osteoarthritis Research Society International (OARSI)~\cite{altman2007atlas} performs independent semi-quantitative 4-scale assessments of osteophytes, joint space narrowing (JSN) and other pathological knee features in different compartments of the joint (\ie, femur, tibia, lateral and medial), as illustrated in Fig.~\ref{fig:oarsi}. Due to the semi-quantitative nature of the KL and OARSI grades, they both suffer from the subjectivity of the raters and hence moderate inter-rater agreement or disagreement. To this end, we aim to develop a computer-aided diagnosis (CADx) method to perform KL and OARSI grading accurately and consistently. 

A few attempts have been made recently to employ convolutional neural networks (CNNs) to assess the knee OA severity from X-rays~\cite{antony2017automatic,tiulpin2018automatic}. They formulate the OA grading problem as an image based classification problem and employ the standard classification CNN models (\eg, ResNet-34). Antony \etal {} formulate OA grading as both classification and regression to train a 5-layer CNN model~\cite{antony2017automatic}. Tiulpin \etal {} employ a 5-layer Siamese network to extract anatomically symmetrical image features from lateral and horizontally-flipped medial patches for grading OA~\cite{tiulpin2018automatic}. These works represent a promising trend in directly adopting deep classification CNNs for OA diagnosis, but fail to integrate some unique properties of the knee OA grading task: (1) KL and OARSI grades measure aspects (joint space, osteophyte) of OA at different granularities (overall, individual knee compartments) that should have underlying coherence. (2) The fine-grained OARSI grades concern more localized visual patterns (\emph{e}.\emph{g}.~, osteophyte), which may not be captured by a standard whole image based classification CNN.


To this end, we propose a novel coherence learning method via keypoint-based pooling network (KPN) to exploit the aforementioned properties for improved OA grading performance. Our approach trains a multi-task neural network to output the KL and 6 OARSI grades simultaneously. We capture the coherence of these grades by estimating their joint distribution using a Gaussian Mixture Model (GMM) trained on expert labeled grades. Given the 7 model-predicted grades, their compliance to the underlying coherence is measured by the multivariate GMM log probability, which forms a novel incoherence loss to train the model on unlabeled data. Furthermore, our novel keypoint-based pooling can extract spatially targeted and pathologically informative convolutional image features to accurately assess the fine-grained OA grades. Specifically, the feature map of the backbone CNN is pooled from a set of automatically detected pre-defined anatomical keypoints along the knee surface, followed by a compartment-aware aggregation to produce features for each OA grade classification task.



Our main contributions are three-fold. We present a novel semi-supervised learning method for OA grading exploiting the underlying coherence of KL and OARSI grades. A keypoint-base pooling scheme is proposed to extract spatially-targeted and pathologically-informative image features for diagnosing OA. We validate that our approach has significant improvements over previous strong baselines for both the KL and OARSI grades, using the Osteoarthritis Initiative (OAI) of 4,796 subjects, multi-center, ten-year observational study.



\section{Methods}
\begin{figure}[t]
\centering
\includegraphics[width=\linewidth]{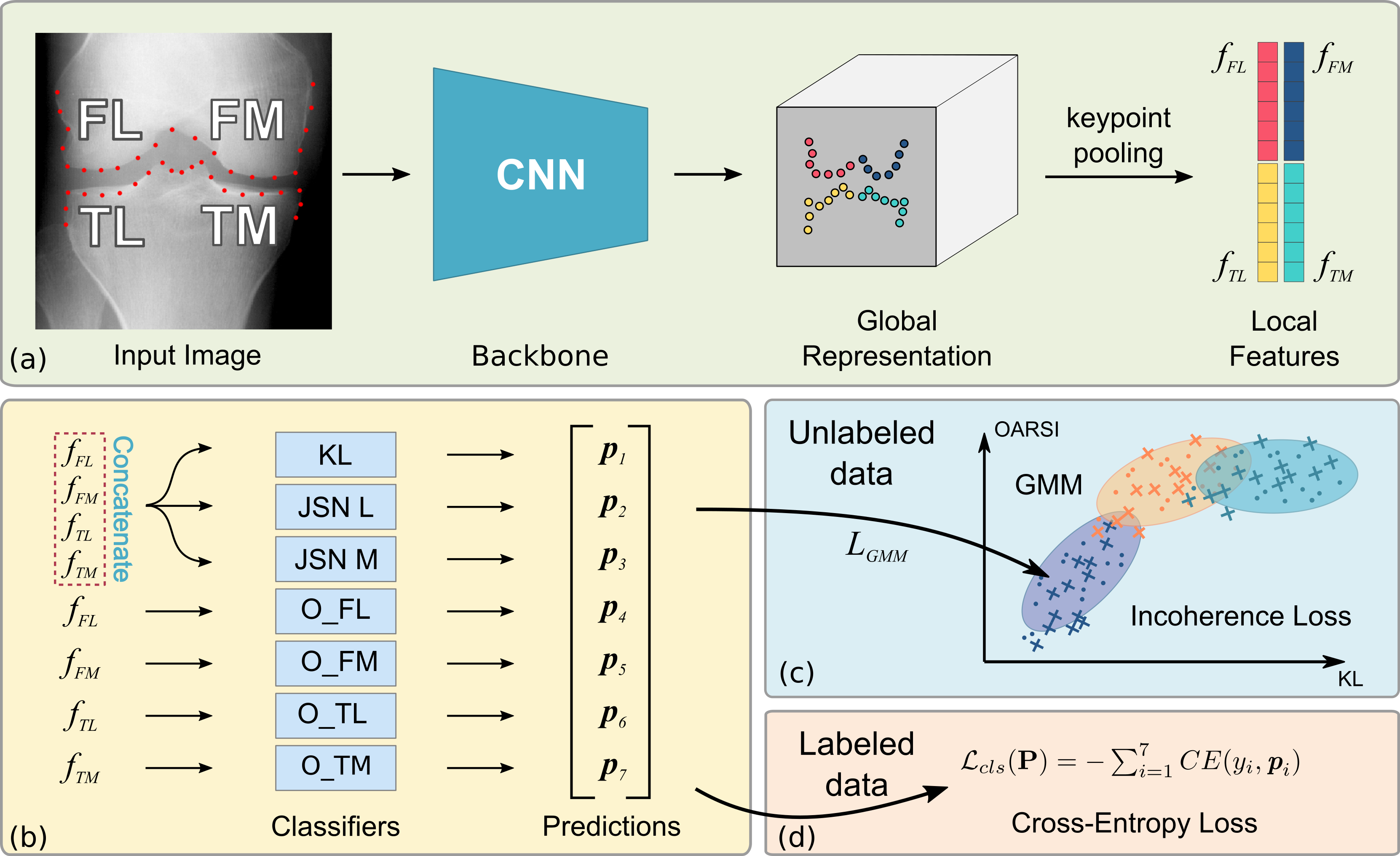}
\caption{Overview of our coherence learning via keypoint-based pooling network (KPN).}
\label{fig:framework}\vspace{-5mm}
\end{figure}

\vspace{-2mm}
Figure~\ref{fig:framework} illustrates the overall framework of the proposed method. We formulate the KL and OARSI grades estimation task as a multi-label multi-class classification problem, modeled by a neural network with 7 classification heads. A novel keypoint-based pooling network is employed to extract CNN features from anatomy driven keypoints for the downstream classification tasks. Our model is trained in a semi-supervised manner incorporating both labeled data and unlabeled data using a novel incoherence loss. We elaborate these components in details as follows.


\vspace{-2mm}
\subsection{Keypoint-based Pooling Network}

The diagnosis of knee OA mostly concerns pathological patterns that are evident along the bone surfaces, such as joint space narrowing and bone spurs.
To mimic this clinical practice, we propose a keypoint-based pooling network to collect CNN features along the bone surfaces, characterized by a set of knee keypoints detected using BoneFinder~\cite{lindner2013fully}, as shown in Figure~\ref{fig:framework}(a). Specifically, we adopt ResNet-50~\cite{he2016deep} as the backbone CNN. Given an input image $I$, we extract the feature map after the 3rd residual block of the backbone ResNet-50, denoted as $\bm{f}$, and detect $N$ keypoints, denoted as $C=\{\mathbf{c}_{1}, \mathbf{c}_{2}, \ldots, \mathbf{c}_{N} \}$. We use bilinear interpolation to pool features $\bm{f}$ at the keypoint coordinates. The resulting features are written as $\bm{f}_{a} = [h(\bm{f}, c_{1}), h(\bm{f}, c_{2}), \ldots, h(\bm{f}, c_{N})]$, where $h(\cdot, \cdot)$ represents bilinear interpolation.

We further split the keypoints into four compartments following the OARSI grades (\ie, FL, FM, TL, TM compartments. T, F, L and M represents tibia, femur, lateral and medial, respectively). The features $\bm{f}_a$ are also splitted into four corresponding parts, \ie $\bm{f}_{FL}$, $\bm{f}_{FM}$, $\bm{f}_{TL}$, and $\bm{f}_{TM}$. Since the KL grade and 2 JSN grades account for the global assessment of the knee, the combined feature $\bm{f}_a$ is used for these three classification tasks. As the 4 osteophyte grades concern bone spurs in the 4 compartments, we use the features from the corresponding compartments individually for the osteophyte grade classification tasks, as illustrated in Figure~\ref{fig:framework}(b). Comparing to the globally pooled feature commonly used in image classification tasks, our keypoint-based pooling in individual compartments provides more spatially targeted and pathologically informative feature to facilitate the downstream classification tasks.

The cross entropy loss is employed to train KPN. We denote the ground-truth grades as a vector $\mathbf{Y}=[ y_{1}, y_{2}, \ldots, y_{7} ]$, where $i=1, 2, \ldots, 7$ indicates the different grade targets (KL, JSN, or osteophyte grade). The predicted probabilities are denoted as $\mathbf{P}=[ \bm{p}_1, \bm{p}_2, \ldots, \bm{p}_7 ]$, where $\bm{p}_i$ is the multi-class softmax probabilities for $i$-th grade target. The classification loss is defined as:
\begin{equation}
    \mathcal{L}_{cls}(\mathbf{P}) = - \sum_{i=1}^{7} CE(y_i, \bm{p}_i),
\end{equation}
where $CE(\cdot, \cdot)$ denotes prevalence weighted cross entropy loss~\cite{rajpurkar2017chexnet}.

\subsection{Coherence Learning with Gaussian Mixture Model}

Since the KL and OARSI grades describe the severity of knee OA from different aspects, an underlying coherence among them should exist by their definitions. For example, a positive KL grade indicates existence of OA-related abnormality, which implies as least one positive OARSI grade. Such coherence can be exploited as a self-supervision to incorporate unlabeled data into training. In particular, we capture the coherence relationship by estimating the joint distribution of the \emph{one-hot representation} of the grades using a GMM~\cite{reynolds2009gaussian}, written as:
\begin{equation}
    \mathbf{P} \sim \sum_{i=1}^{K} \alpha_{i} \mathcal{N} (\boldsymbol{\mu}_{i}, \boldsymbol{\Sigma}_{i}) ,
\end{equation}
where $K$ is the number of Gaussians, $\alpha_{i}$, $\boldsymbol{\mu}_i \in {\mathrm{R}}^{D} $ and $\boldsymbol{\Sigma}_i \in {\mathrm{R}}^{D \times D} $ are the weight, mean and covariance matrix for the $i$-th Gaussian, respectively. $D$ is the length of the one-hot representation. Since the KL grade has 5 scales (0 to 4) and OARSI grades have 4 scales (0 to 3), the length of the one-hot representation is $D=5 + 4 \times 6 = 29$. We build the GMM based on the one-hot representation so that we can directly perform inference on the multi-label multi-class softmax probabilities $\mathbf{P}$ produced by the KPN. The GMM is pre-trained on the expert labeled KL and OARSI grades using the classic Expectation Maximization method~\cite{moon1996expectation}. Since the softmax probabilities produced by the KPN are not discrete $\{0, 1\}$, a zero-mean Gaussian perturbation with a standard deviation $\sigma$ is added to the one-hot representation when training the GMM.

Given the model predicted softmax probabilities $\mathbf{P}$ of the KL and OARSI grades, we measure their coherence using the multivariate log probability of $\mathbf{P}$ in the GMM:
\begin{equation}
    g(\mathbf{P})=\log \left ( \sum_{i=1}^{K} \alpha_{i} \frac{ \exp \left(-\frac{1}{2} (\mathbf{P}-\mu_i) \boldsymbol{\Sigma}_i ^ {-1} (\mathbf{P}-\mu_i)^T \right )}{\sqrt{(2 \pi) ^ D } | \boldsymbol{\Sigma}_i | } \right ), 
\end{equation}
where $| \cdot |$ denotes matrix determinant. Higher $g(\mathbf{P})$ indicates that the predicted grades are more likely to be coherent. Therefore, we proposed a novel \emph{incoherence loss} based on $g(\mathbf{P})$, with a non-linear mapping to bring it into $[0, \infty]$ range:
\begin{equation}
    \mathcal{L}_{gmm}(\mathbf{P}) = \frac{1}{\tau} \log \left (1 + e^{- \tau \cdot g(\mathbf{P})} \right ),
\end{equation}
where $\tau$ is a temperature hyper-parameter to account for the scale of $g(\mathbf{P})$, partially inspired by the InfoNCE loss used in contrastive learning~\cite{oord2018representation}. Since the incoherence loss does not rely on ground truth grades, it can be used to incorporate a large number of unlabeled data to facilitate model training. Our method is trained in a semi-supervised manner using a weighted combination of the classification loss and the incoherence loss:
\begin{equation}
    \mathcal{L} = \lambda_{cls} \mathcal{L}_{cls} + \lambda_{gmm} \mathcal{L}_{gmm},
\end{equation}
where $\lambda_{cls}$ and $\lambda_{gmm}$ are the weights for balancing the two losses. While the classification loss is calculated only on labeled data, the incoherence loss is calculated on all data including the unlabeled ones. Since the model predicts random, completely incoherent grades at the beginning of training, it may lead to a large incoherence loss that throws off the training. To mitigate this behavior, we start the training with 1 epoch warm-up training phase using only the classification loss, and the incoherent loss kicks in after the warm-up phase.

\section{Experimental Results}

\noindent\textbf{Experiment Setup.}
We evaluate our method on the publicly available OAI dataset, which is a multi-center, longitudinal, prospective observational study of knee OA on 4,796 subjects. We extracted 23,249 labeled knee X-ray images with KL and OARSI grades labeled by experts and 21,763 images without OA grade labels. In the labeled images, the distribution of KL grades are shown in Table~\ref{tab:oai_distrib}.
In our experiment, we conduct a subject-stratified random split of the images with 2:1:1 ratio into training, validation and testing sets, respectively.

\begin{table}[]
\centering
\caption{Distribution of KL and OARSI grades of OAI dataset.}
\label{tab:oai_distrib}
\begin{tabular}{c|c|cccccc}
\hline
\multirow{2}{*}{Grade} & \multirow{2}{*}{KL} & \multicolumn{2}{c}{FO} & \multicolumn{2}{c}{TO} & \multicolumn{2}{c}{JSN} \\ \cline{3-8}
                       &                     & L          & M         & L          & M         & L          & M          \\ \hline

0                      & 2827                & 13646      & 11728     & 13976      & 8043      & 20207      & 10676      \\
1                      & 2930                & 5554       & 5319      & 6124       & 10870     & 1337       & 6884       \\
2                      & 10231               & 2045       & 2486      & 1373       & 2552      & 1197       & 4506       \\
3                      & 5578                & 2004       & 3716      & 1776       & 1784      & 508        & 1183       \\
4                      & 1683                & -          & -         & -          & -         & -          & - \\ \hline
\end{tabular}
\end{table}

We measure the performance of all evaluated method using the \textit{accuracy} and \textit{root mean squared error (RMSE)} of the KL grade and 6 OARSI grades, including JSN grades in 2 compartments and osteophyte grades in 4 compartments.

\vspace{2mm}
\noindent\textbf{Implementation Details.}
Knee ROIs are extracted from the original X-ray images centered at the knee joint center based on the keypoints detected by BoneFinder. Left knees are flipped so that all knees have the same orientation. The ROIs are resized to $320\times 320$ pixels, and augmented with random $[-10, 10]$ pixels translation, $[-5, 5]$ degrees rotations and $[0.95, 1.05]$ scales. The hyper-parameters of the proposed method are: $K=10$, $\lambda_{cls}=1.0$, $\lambda_{gmm}=1.0$, $\tau=0.5$, $\sigma=0.2$. All networks are trained for 60 epochs using Adam optimizer with a batch size of 48, a weight decay of $10^{-2}$ and an initial learning rate of $10^{-4}$. The learning rate decays by $0.5$ after 10 and 30 epochs. 

\vspace{2mm}
\noindent\textbf{Comparison with Baseline Methods.}
We compare our method with several strong classification baseline methods, \textit{ResNet-18}, \textit{ResNet-34} and \textit{ResNet-50}. The baseline methods are trained using only the labeled data, while our method also exploits the unlabeled data. Since the authors of \cite{tiulpin2018automatic} conclude based on experiments that their method performs similarly to ResNet-34 (reporting 67.49\% KL grade accuracy in their experiment setting), we do not evaluate it in addition to ResNet-34. Another related method \cite{antony2017automatic} reports 60.3\% KL grade accuracy, which clearly falls behind strong classification baselines like ResNet, as demonstrated in our experiments.

Table~\ref{tab:sota} summarizes the experiment results. ResNet-50 performs overall the best among the three baseline methods, reporting a KL grade accuracy of 70.8\% and an average OARSI grade accuracy of 75.3\%. Our method, using ResNet-50 as backbone, significantly outperforms ResNet-50 on the KL grade, measuring an accuracy of 72.7\%. Our method also results in the best performance on the average OARSI grade as well as all individual OARSI grades, except on O\_TL (\ie the osteophyte grade in the tibia lateral compartment). Although the accuracy of O\_TL degrades slightly, the RMSE is improved, indicating that our method reduces large errors in O\_TL grade. We stress that since the KL and OARSI grades are semi-quantitative and observer subjective, there is an inherent accuracy ceiling when comparing against human labeled ground truth. An accuracy above 70\% may already indicate human-level performance~\cite{tiulpin2018automatic}.


\newcommand{\STAB}[1]{\begin{tabular}{@{}c@{}}#1\end{tabular}} 
\setlength{\tabcolsep}{1.5mm}
\begin{table}[t]
\centering
\caption{Summary of knee OA grade estimation results in comparison to baseline methods. O indicate osteophyte grade. L and M indicate lateral and medial, respectively. F and T indicate femur and tibia, respectively.}
\begin{tabular}{l|l|c|ccccccc} \hline
& \multirow{2}{*}{Method} & \multirow{2}{*}{KL}    & \multicolumn{7}{c}{OARSI}  \\ \cline{4-10}
&           &       & JSN\_L & JSN\_M & O\_FL & O\_FM & O\_TL & O\_TM & Average  \\ \hline
\multirow{4}{*}{\STAB{\rotatebox[origin=c]{90}{Accuracy}}} 
& ResNet-18 & 0.691 & 0.929  & 0.767  & 0.645 & 0.673 & 0.719 & 0.672 & 0.734    \\
& ResNet-34 & 0.703 & 0.926  & 0.767  & 0.655 & 0.702 & 0.741 & 0.694 & 0.748    \\
& ResNet-50 & 0.708 & 0.924  & 0.773  & 0.680 & 0.697 & \bf 0.750 & 0.693 & 0.753    \\
& Proposed     & \bf 0.727 & \bf 0.938  & \bf 0.792  & \bf 0.686 & \bf 0.712 & 0.746 & \bf 0.712 & \bf 0.764    \\ \hline
\multirow{4}{*}{\STAB{\rotatebox[origin=c]{90}{RMSE}}} 
& ResNet-18 & 0.695 & 0.286  & 0.499  & 0.798 & 0.869 & 0.626 & 0.631 & 0.618    \\
& ResNet-34 & 0.672 & 0.298  & 0.500  & 0.770 & 0.785 & 0.600 & 0.608 & 0.593    \\
& ResNet-50 & 0.683 & 0.321  & 0.491  & 0.747 & 0.768 & 0.579 & 0.604 & 0.585    \\
& Proposed     & \bf 0.639 & \bf 0.278  & \bf 0.472  & \bf 0.731 & \bf 0.719 & \bf 0.571 & \bf 0.597 & \bf 0.561     \\ \hline
\multirow{4}{*}{\STAB{\rotatebox[origin=c]{90}{Kappa}}}     
& ResNet-18 & 0.802 & 0.893  & 0.857  & 0.676 & 0.763 & 0.791 & 0.758 & 0.776 \\
& ResNet-34 & 0.792 & 0.899  & 0.857  & 0.678 & 0.719 & 0.766 & 0.738 & 0.780 \\
& ResNet-50 & 0.799 & 0.880  & 0.863  & 0.701 & 0.762 & 0.800 & 0.767 & 0.795 \\
& Proposed     & \bf 0.825 & \bf 0.906  & \bf 0.873  & \bf 0.721 & \bf 0.789 & \bf 0.810 & \bf 0.773 & \bf 0.812
\\ \hline
\end{tabular}
\vspace{-2mm}
\label{tab:sota}
\end{table}

\vspace{2mm}
\noindent\textbf{Ablation Study.}
We conduct an ablation study to further analyze both the individual and compound effects of the proposed KPN and coherence learning, summarized in Table~\ref{tab:ablation}. All methods evaluated in the ablation study use ResNet-50 as the backbone network. When KPN is not employed, the feature produced by global average pooling of the last feature map of ResNet-50 is used for all downstream OA grade classification tasks.

Comparing to the baseline, employing KPN leads to insignificant degradation in the KL grade accuracy (\ie from 0.708 to 0.703) but notable improvement in the average OARSI grade accuracy (\ie from 0.752 to 0.761), respectively. RMSEs on both the KL and OARSI grades are improved. These results show that while the keypoint-based pooling helps to improve the fine-grained OARSI grades, its effect on the composite KL grade is limited, especially measured by the accuracy. We conjugate that the improvements on the fine-grained OARSI grades are owning to the spatially targeted and pathologically informative features produced by KPN, where the visual features along different compartments of joint surface are pooled for classifying the corresponding OARSI grades.

On the other hand, employing the coherence learning leads to significant improvement in the KL grade performance (\ie accuracy from 0.708 to 0.723, RMSE from 0.683 to 0.639) but insignificant change in the average OARSI grade performance. These results suggest that when using the standard ResNet-50, imposing the OA grade coherence constraint in training has more positive effect on the composite KL grade than the fine-grained OARSI grades. We posit that while the globally pooled feature in ResNet-50 may be sufficient for assessing the composite KL grade, it lacks spatially targeted representation needed for the fine-grained OARSI grades. This limitation prevents the grade incoherence loss to be effectively propagated to the OARSI grades. 

By employing both the KPN and coherence learning, our method achieves significantly improved performances on both the KL and OARSI grades, comparing to all three evaluated variations. The comparison demonstrates the contributions and necessities of the two novelties in our method in achieving the optimal knee OA grading performance. 

\setlength{\tabcolsep}{1.5mm}
\begin{table}[t]
\centering
\caption{Ablation study of the proposed KPN and coherence learning (CL). The best performance is marked in bold.}
\begin{tabular}{c|c|cc|cc} \hline
\multirow{2}{*}{KPN} & \multirow{2}{*}{CL} & \multicolumn{2}{c|}{KL}           & \multicolumn{2}{c}{OARSI average}  \\ \cline{3-6}
                           &                      & Accuracy  & RMSE      & Accuracy  & RMSE      \\ \hline
                           &                      & 0.708     & 0.683     & 0.752     & 0.585     \\
\checkmark                 &                      & 0.705     & 0.668     & 0.761     & 0.563     \\
                           & \checkmark           & 0.723     & 0.664     & 0.750     & 0.587     \\
\checkmark                 & \checkmark           & \bf 0.727 & \bf 0.639 & \bf 0.764 & \bf 0.561    
\\ \hline
\end{tabular}
\label{tab:ablation}
\end{table}

\vspace{2mm}
\noindent\textbf{Analysis on Various Levels of Demand for Labeled (Unlabeled) Data.}
A major contribution of this work is the coherence learning, which enables our method to incorporate unlabeled data during training and hypothetically be less demanding on the amount of labeled data. To analyze the effectiveness of the coherence learning in this aspect, we conduct an experiment to evaluate the performances of the baseline ResNet-50 and our method trained with 20\%, 40\% and 100\% of the labeled data, respectively. Table~\ref{tab:lessdata} summarizes the results of this experiment. We observe that the baseline ResNet-50 performs poorly when only 20\% of labeled data are used, and gradually improve as more labeled data are used. In contrast, using only 20\% labeled data, our method competes head-to-head with ResNet-50 using 100\% labeled data. When trained using 40\% labeled data, our method significantly outperforms ResNet-50 using 100\% labeled data. Note that training our method using 100\% labeled data leads to only marginal improvement over using 40\% labeled data, indicating that our method is less demanding on labeled data.

\setlength{\tabcolsep}{1mm}
\begin{table}[t]
\centering
\caption{Performances of ResNet-50 and the proposed method when trained with 20\%, 40\% and 100\% labeled data.}
\begin{tabular}{c|cc|cc|cc|cc} \hline
Labeled & \multicolumn{4}{c|}{KL}                                   & \multicolumn{4}{c}{OARSI average}                         \\ \cline{2-9}
data    & \multicolumn{2}{c|}{ResNet-50} & \multicolumn{2}{c|}{Proposed} & \multicolumn{2}{c|}{ResNet-50} & \multicolumn{2}{c}{Proposed}  \\ \cline{2-9}
used    & Accuracy & RMSE               & Accuracy & RMSE          & Accuracy & RMSE               & Accuracy & RMSE           \\ \hline
20\%      & 0.662    & 0.715              & 0.684    & 0.677         & 0.690    & 0.661              & 0.756    & 0.587          \\
40\%      & 0.695    & 0.698              & 0.725    & 0.642         & 0.730    & 0.642              & 0.765    & 0.570          \\
100\%     & 0.708    & 0.683              & 0.727    & 0.639         & 0.752    & 0.585              & 0.764    & 0.561    
\\ \hline
\end{tabular}
\vspace{-2mm}
\label{tab:lessdata}
\end{table}

\section{Conclusion \& Future Work}
We proposed a new method for computer-aided diagnosis (CAD) of knee OA in X-ray images: comprising two major novelties: 1) a coherence learning scheme that exploits the underlying coherence of OA grade to enable learning from unlabeled data, and 2) a keypoint-based pooling network to extract spatially targeted and pathologically informative features for OA diagnosis.
We discovered that: 1) the diagnosis of knee OA in fine-grain level, specifically joint space narrowing and osteophytes, can be improved by extracting better pathological features via keypoint-based pooling over image features, and 2) there is an implicit coherence among KL and OARSI grades, which can be exploited to improve the training of knee OA CAD models.
Our method is validated on the OAI dataset demonstrating significant improvements over the state-of-the-art methods on both the KL and OARSI grading tasks. Individualized contributions of the proposed KPN and coherence learning are analyzed and justified in our ablation study. The proposed method is also applicable to grading arthritis in other joints that share the same or similar physiology/anatomy nature (\eg, hand, hip, etc.). This will be investigated in future work.
Furthermore, our coherence learning can be utilized in a generic multi-task learning setting such as classification/detection of multiple attributes.


\bibliographystyle{splncs04}
\bibliography{miccai2020}

\end{document}


\newcommand{\STAB}[1]{\begin{tabular}{@{}c@{}}#1\end{tabular}} 
\setlength{\tabcolsep}{1.5mm}
\begin{table}
\centering
\caption{\textbf{Evaluation of the temperature parameter $\tau$'s influence on the performance of the proposed method.} Our method with different $\tau$ values always outperforms the baseline ResNet-50, although to different extents.}
\begin{tabular}{l|l|c|ccccccc} \hline
& \multirow{2}{*}{Parameter} & \multirow{2}{*}{KL}    & \multicolumn{7}{c}{OARSI}  \\ \cline{4-10}
&           &       & JSN\_L & JSN\_M & O\_FL & O\_FM & O\_TL & O\_TM & Average  \\ \hline \hline
\multirow{6}{*}{\STAB{\rotatebox[origin=c]{90}{Accuracy}}} 
& ResNet-50   & 0.708 & 0.924 & 0.773 & 0.680 & 0.697 & 0.750 & 0.693 & 0.753    \\ \cline{2-10}
& $\tau=0.25$ & 0.721 & 0.941 & 0.790 & 0.699 & 0.729 & 0.770 & 0.714 & 0.774    \\
& $\tau=0.5$  & 0.727 & 0.938 & 0.792 & 0.686 & 0.712 & 0.746 & 0.712 & 0.764    \\
& $\tau=0.75$ & 0.734 & 0.936 & 0.799 & 0.696 & 0.719 & 0.759 & 0.708 & 0.770    \\
& $\tau=1.0$  & 0.721 & 0.937 & 0.796 & 0.687 & 0.715 & 0.760 & 0.705 & 0.766    \\
& $\tau=1.5$  & 0.719 & 0.940 & 0.796 & 0.692 & 0.727 & 0.759 & 0.704 & 0.770    \\ \hline \hline
\multirow{6}{*}{\STAB{\rotatebox[origin=c]{90}{RMSE}}} 
& ResNet-50   & 0.683 & 0.321 & 0.491 & 0.747 & 0.768 & 0.579 & 0.604 & 0.585    \\
\cline{2-10}
& $\tau=0.25$ & 0.663 & 0.271 & 0.472 & 0.728 & 0.725 & 0.554 & 0.578 & 0.555    \\
& $\tau=0.5$  & 0.639 & 0.278 & 0.472 & 0.731 & 0.719 & 0.571 & 0.597 & 0.561    \\
& $\tau=0.75$ & 0.622 & 0.273 & 0.465 & 0.736 & 0.717 & 0.563 & 0.579 & 0.556    \\
& $\tau=1.0$  & 0.628 & 0.268 & 0.463 & 0.762 & 0.734 & 0.565 & 0.591 & 0.564    \\
& $\tau=1.5$  & 0.653 & 0.271 & 0.462 & 0.728 & 0.730 & 0.566 & 0.597 & 0.559    \\ \hline
\end{tabular}
\vspace{8mm}
\label{tab:tau_comparison}

\setlength{\tabcolsep}{1.5mm}
\begin{threeparttable}
\caption{
\textbf{Effect of employing an auxiliary regression head.} Antony \etal{}~[2] trained a 5-layer classification CNN with an auxiliary regression head. Although poorer results were reported in [2], the auxiliary regression head is an orthogonal novelty that can be added to the baseline ResNet-50 (R-50) and our method. 
%
}
\begin{tabular}{l|l|c|ccccccc} \hline
& \multirow{2}{*}{Method} & \multirow{2}{*}{KL}    & \multicolumn{7}{c}{OARSI}  \\ \cline{4-10}
&           &       & JSN\_L & JSN\_M & O\_FL & O\_FM & O\_TL & O\_TM & Average  \\ \hline
\multirow{5}{*}{\STAB{\rotatebox[origin=c]{90}{Accuracy}}} 
& Original [2]* & 0.603 & - & - & - & - & - & - & -    \\
& R-50 & 0.708 & 0.924  & 0.773  & 0.680 & 0.697 & 0.750 & 0.693 & 0.753    \\
& R-50 w. [2] & 0.708 & \bf 0.938 & 0.765 & 0.677 & 0.702 & \bf 0.761 & 0.699 & 0.757   \\
& Prop. & \bf 0.727 & \bf 0.938  & 0.792  & \bf 0.686 & 0.712 & 0.746 & 0.712 & 0.764    \\
& Prop. w. [2] & 0.721 & \bf 0.938 & \bf 0.795 & 0.685 & \bf 0.723 & 0.757 & \bf 0.723 & \bf 0.770   \\ \hline
\multirow{5}{*}{\STAB{\rotatebox[origin=c]{90}{RMSE}}} 
& Original [2]* & 0.898 & - & - & - & - & - & - & -    \\
& R-50 & 0.683 & 0.321 & 0.491 & 0.747 & 0.768 & 0.579 & 0.604 & 0.585    \\
& R-50 w. [2] & 0.676 & 0.263 & 0.498 & 0.731 & 0.772 & \bf 0.554 & 0.583 & 0.567    \\
& Prop. & 0.639 & 0.278 & 0.472 & 0.731 & 0.719 & 0.571 & 0.597 & 0.561    \\
& Prop. w. [2] & \bf 0.629 & \bf 0.258 & \bf 0.463 & \bf 0.726 & \bf 0.716 & \bf 0.554 & \bf 0.560 & \bf 0.546    \\ \hline
\end{tabular}
\begin{tablenotes}
\item *Numbers are from the original paper, which were obtained under different experiment settings. Only the KL grade is estimated in [2].
\end{tablenotes}
\end{threeparttable}
\label{tab:reg_loss_comparison}

\end{table}

\begin{figure}
\centering
\includegraphics[width=0.9\linewidth]{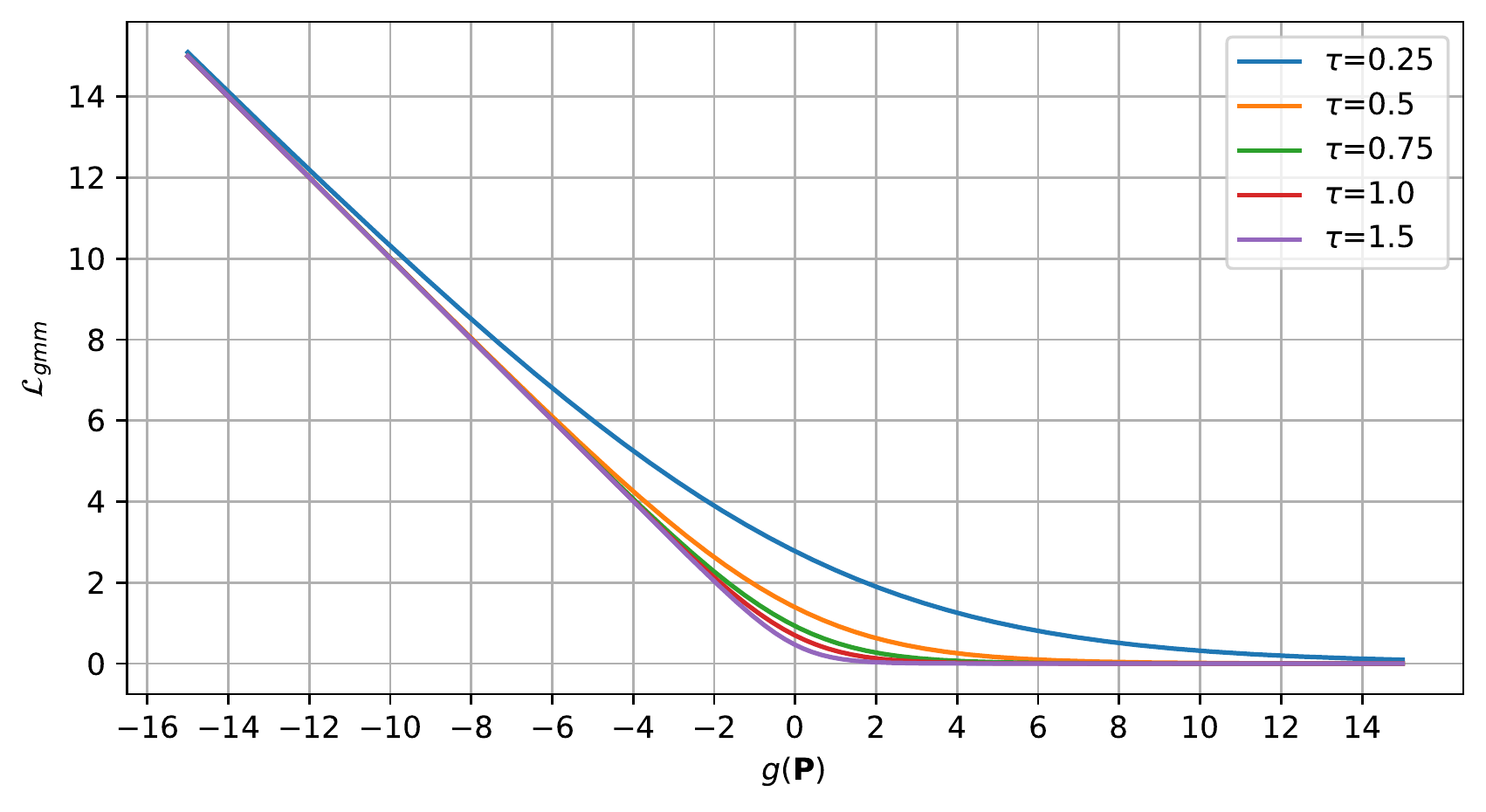}
\caption{Visualization of the non-linear mapping in Equation (4) under different $\tau$ values. \textbf{[x-axis]} GMM log-probability. \textbf{[y-axis]} Incoherence loss.}
\vspace{8mm}
\centering
\includegraphics[width=0.9\linewidth]{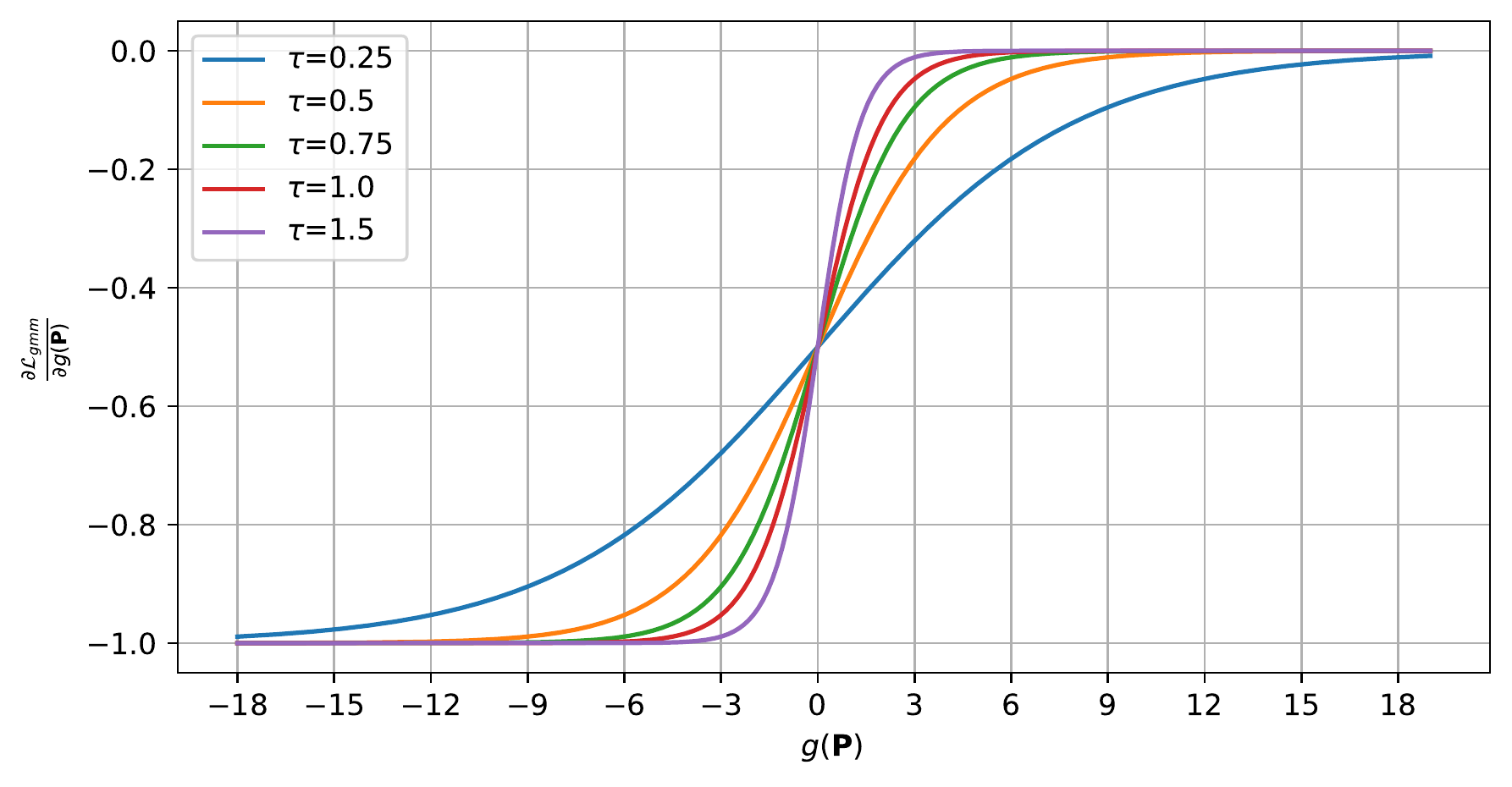}
\caption{Visualization of gradient of the non-linear mapping in Equation (4) under different $\tau$ values. \textbf{[x-axis]} GMM log-probability. \textbf{[y-axis]} Gradient of the incoherence loss.}
\end{figure}